\documentclass[twocolumn]{aastex62}
\usepackage{natbib}
\usepackage{cancel} 

\usepackage{graphicx}
\usepackage[space]{grffile}
\usepackage{latexsym}
\usepackage{amsfonts,amsmath,amssymb}
\usepackage{url}
\usepackage[utf8]{inputenc}
\usepackage{fancyref}
\usepackage{hyperref} 
\usepackage{breakurl}

\usepackage{ulem} 
\usepackage{color} 

\usepackage{pifont}
\newcommand{\cmark}{\ding{51}}%
\newcommand{\xmark}{\ding{55}}%

\begin{document}

\title{The Intrinsic Stochasticity of the $^{56}$Ni Distribution of Single-Degenerate Type Ia Supernovae}


\author{Chris Byrohl}
\affiliation{Max-Planck-Institut f\"ur Astrophysik, Karl-Schwarzschild-Stra\ss e 1, 85748 Garching, Germany}
\author{Robert Fisher}
\affiliation{Department of Physics, University of Massachusetts Dartmouth, 285 Old Westport Road, North Dartmouth, Ma. 02740, USA}
\author{Dean Townsley}
\affiliation{Department of Physics \& Astronomy, Box 870324, University of Alabama, Tuscaloosa, AL. 35487-0324, USA}



\begin{abstract}

Binary Chandrasekhar-mass white dwarfs accreting mass from non-degenerate
stellar companions through the single-degenerate channel have reigned for
decades as the leading explanation of Type Ia supernovae. 
Yet, a comprehensive theoretical explanation has not yet emerged to explain the
expected properties of the canonical near-Chandrasekhar-mass white dwarf model.
A  simmering phase within the convective core of the white dwarf leads
to the ignition of one or more flame bubbles scattered across the core.
Consequently, near-Chandrasekhar-mass single-degenerate SNe Ia are inherently
stochastic, and are expected to lead to a range of outcomes,
from subluminous SN 2002cx-like events, to overluminous SN 1991T-like events. 
However, all prior simulations of the single-degenerate channel carried
through the detonation phase have set the ignition points as free parameters.
In this work, for the first time, we place ignition points as predicted by {\it ab initio} models 
of the convective phase leading up to ignition, and follow through the detonation
phase in fully three-dimensional simulations. Single-degenerates in this framework
are characteristically overluminous. Using a 
statistical approach, we determine
the $^{56}$Ni mass distribution arising from stochastic ignition.
While there is a total spread of $\gtrsim 0.2 M_{\odot}$ for detonating models,
the distribution is strongly left-skewed, and with a narrow standard deviation of
$\simeq 0.03 M_{\odot}$. 
Conversely, if single-degenerates are not overluminous but primarily
yield normal or failed events, then the models require fine-tuning of the ignition
parameters, or otherwise require revised physics or progenitor models. We
discuss implications of our findings for the modeling of single-degenerate SNe
Ia.

\end{abstract}

\keywords{supernovae: general --- supernovae: individual (1991T, 2002cx) --- hydrodynamics --- white dwarfs}

\section{Introduction}

Numerous classic works identified white dwarfs accreting to near the
Chandrasekhar mass $M_{\rm ch}$ in binary systems as candidate progenitors of Type Ia
supernovae (SNe Ia) -- e.g. \citet {arnett69}, \citet {Whelan_Iben_1973}, and
\citet {nomotoetal84}. This classic picture was long thought to provide an
explanation for the uniformity of brightnesses observed in SNe Ia \citep
{phillips93}.

The nature of the dominant production channel for SNe Ia has long been unclear
\citep{BranchSearchProgenitorsType1995} and more recently, the classic picture
of near $M_{\rm ch}$ progenitors has been substantially revised, with single-degenerates
now widely believed to be rare in nature. The single-degenerate channel has been
shown to be inconsistent with a range of constraints, including the delay-time
distribution, the absence of hydrogen, and the absence of companions
\citep{maozetal14}. Single-degenerates are also inconsistent with observational
and theoretical rate predictions \citep {maozmannucci12}. However, recent
observations have provided strong evidence
that Chandrasekhar-mass white dwarf SNe Ia do occur in at least some systems in
nature. Hard X-ray spectra of the 3C 397 supernova remnant (SNR) are consistent
with electron captures which arise during nuclear burning at high densities
typical of Chandrasekhar-mass 
white dwarfs \citep{yamaguchietal14, yamaguchietal15}. Additional X-ray and
infrared observations of the Kepler SNR suggest that it was an overluminous
single-degenerate supernova \citep {katsudaetal15}. Furthermore, the 
pre-maximum light shock signature detected in both a 
subluminous SN Ia 2012cg  \citep{marionetal16} and a normal SN Ia iPTF14atg  \citep{caoetal15}
similar to theoretical predictions
of the shock interaction with the companion star \citep {kasen10}, although these 
observations have also been contested \citep {kromeretal16, shappeeetal18}.


A large body of theoretical and computational work has explored possible 
mechanisms for single-degenerate SNe Ia \citep{maozmannucci12}.
Many single-degenerate explosion mechanisms begin with a deflagration in the
convective core of a near-$M_{\rm Ch}$ WD \citep{nomotoetal84}.
From this common starting point, authors have explored the possibility
of pure deflagrations \citep{ropkeetal07b, jordanetal12a, kromeretal13},
deflagration-to-detonation transitions (DDTs) (\citet{khokhlov91, ropkeetal07a,
  Seitenzahl3Dddt2013,maloneetal14,Martinez-Rodriguez2017,daveetal17} and many more,
and gravitationally-confined detonations 
(GCDs) \citep {plewaetal04, ropkeetal07a, townsleyetal07,jordanetal08,meakinetal09,seitenzahletal16}. 
The viability of the proposed explosion mechanisms hinges crucially 
on the nature of the flame ignition during the convective phase. In particular,
the GCD mechanism relies upon an offset ignition to buoyantly drive the 
flame bubble through breakout. Because the vigor of the GCD mechanism relies
upon maintaining the WD intact until the ash collides
at a point opposite of breakout, its viability is diminished as the
ignitions become more centrally concentrated and multi-point.
In contrast, a pure deflagration model produces good agreement with 
observations of the subclass of SNe Iax \citep {kromeretal13},
but requires a vigorous deflagration phase with several simultaneous 
near-central ignitions. Both the pure deflagration and the GCD mechanism
require that the flame surface does not undergo a transition to 
a detonation prior to breakout, as the DDT model does. Furthermore, 
there exists the possibility that a detonation does not arise during the 
initial ash collision subsequent to breakout, and that the WD
remains gravitationally bound, leading to a subsequent contraction
and a detonation through the pulsationally-assisted GCD (PGCD)
mechanism \citep {garciasenzbravo05, jordanetal12a}.

Because the ignition of a flame bubble in the convective core of the
white dwarf is inherently stochastic, outcomes ranging from subluminous 
through overluminous SNe Ia are expected to arise in Chandrasekhar-mass SNe Ia. 
The ignition arises within a highly-turbulent (Reynolds number Re $\sim 10^{15}$) 
convective flow \citep {isernetal17}, with the detailed outcome critically dependent upon the 
high-end tail of the temperature distribution. For many years, the distribution
of ignition points was poorly constrained by theory and 
simulation \citep{garciasenzwoosley95, woosleyetal04}. Early studies suggested
multi-point ignitions as a viable scenario, which has only been revised recently
as it became possible to begin to simulate these
crucial last minutes of the simmering phase in full 3D simulations.
For example, \citet{zingaleetal11} and \citet{nonakaetal11} performed a numerical study for 
a WD with a central density $2.2\times 10^{9}$ g cm$^{-3}$ and central temperature $6.25\times 10^{8}$ K
to determine the probability distribution of hot spots triggering the deflagration phase. 
\citet {zingaleetal11} demonstrated that most ignitions for the
progenitor considered occur at a single point at radial offsets below 100 km from the
center, and most likely at about 50~km. Consequently, these {\it ab initio}
simulations point towards a low amount of deflagration energy resulting from
a small single bubble, buoyancy-driven ignition, in contrast to prior simulations
which often invoked multiple-bubble ignitions.
It has been known for some time that such low-deflagration energies generally
lead to large amounts of
\citet{ropkeetal07a} run a series of simulations with off-centered ignitions
demonstrating an anti-correlation of deflagration yield and ignition offset.
However, initial offsets do not include ignitions below 50~km, thus neglecting
roughly half of the ignitions expected from results in \citet{nonakaetal11}.
\citet{hillebrandtetal07} propose that off-centered, lobsided explosions, such
as those following the deflagration phase simulated in \citet{ropkeetal07a},
might explain overluminous SN Ia events.

Recent theoretical work explored the physics of stochastic ignition close to the
WD's center
using semi-analytic methods in-depth, and demonstrated that single-bubble ignitions are
generally buoyancy-dominated, leading to a weak deflagration phase 
\citep{fisherjumper15}. Consequently, as \citet{fisherjumper15} argued,
single-bubble ignitions tend to lead to the production of a relatively large
amount of $^{56}$Ni and hence an overluminous SN Ia. This theoretical work 
was soon given observational support when spectral modeling of the nebular 
phase of SNe Ia revealed the canonical bright event SN 1991T had an inferred ejecta
mass of 1.4 $M_{\odot}$ \citep{childressetal15}. 
Most recently, \citet {jiangetal18} have examined the early-phase light curves of 40 SNe Ia in the optical, UV, and NUV, and demonstrated that all six luminous 91T- and 99aa-like events in their sample are associated with an 
early-excess consistent with a $^{56}$Ni-abundant outer layer, as expected in the GCD scenario.
Subsequent three-dimensional simulations
of a buoyantly-driven single bubble ignition confirmed a large amount of $^{56}$Ni 
consistent with SN 1991T \citep {seitenzahletal16}. However, because the stable
IGEs tend to be buoyantly-driven in the GCD model, \citet {seitenzahletal16} found 
that the observed stable IGEs at low velocities in their model could only be reproduced 
along a line of sight centered around the detonation region. While there are
systemic differences in how the LEAFS code used by \citet {seitenzahletal16}
treats subgrid scale turbulent nuclear burning in comparison to FLASH  -- see
e.g. \citet {jordanetal08}, it is possible that the bulk of this inconsistency
could be rectified by a DDT as opposed to a GCD model. In particular, \citet {fisherjumper15} 
noted that buoyantly-driven ignitions will lead to a large amount of
$^{56}$Ni and an overluminous SNe Ia in both the DDT and GCD models.

This recent observational and theoretical progress motivates the current study,
in which we explore the inherent stochasticity of near-Chandrasekhar mass white dwarfs in
the single-degenerate channel, from ignition through detonation. 
In Section~\ref{sec:methodology}, we
shortly summarize the simulation setup and the assumed initial hot spot
distribution. In Section~\ref{sec:results}, we describe the WD's evolution from
ignition to its possible detonation depending on the ignition's offset to the
center of mass and link our findings to the initial hot spot distribution. In
Section~\ref{sec:discussion}, we discuss possible uncertainties in our modeling
before summarizing our findings in Section~\ref{sec:conclusions}.

\section{Methodology}
\label{sec:methodology}

Our simulations were performed with the 3D Eulerian adaptive mesh refinement 
(AMR) code FLASH 4.3 \citep{Fryxell_2000} solving the hydrodynamic equation
with the directionally split piecewise-parabolic method (PPM).  We use a 
tabular Helmholtz equation of state taking into account radiation, nuclei,
electrons, positrons and corrections for Coulomb effects, which remains valid in
the electron degenerate relativistic regime \citep{Timmes_2000}.  Flame physics is 
modeled by an advection-diffusion-reaction equation.  Nuclear energy generation
is incorporated using a simplified treatment of the flame energetics 
\citep{townsleyetal07, townsleyetal09, townsleyetal16}.  Self-gravity is accounted for by 
a multipole solver \citep{Couch_2013}  up to order $l=6$ with isolated 
boundary conditions.

The progenitor model of the white dwarf used assumes a mass of 
$1.38\ M_{\odot}$ and a uniform 50/50 carbon/oxygen (C/O) composition. 
See Section~\ref{sec:discussion}
for a  discussion of the impact of non-zero stellar progenitor metallicity.
The white dwarf has a central temperature stratification, 
including an adiabatic core with central density $2.2 \times 10^9$ g cm$^{-3}$
and  temperature $7 \times 10^8$ K, pressure-matched onto an isothermal
envelope with temperature $10^7$ K \citep {jacksonetal10, kruegeretal12}.
Furthermore, the central density of our WD progenitor 
is a standard value commonly considered in the literature, because 
higher-central density WD progenitors produce anomalously high abundances of 
Fe-peak elements, including $^{48}$Ca, $^{54}$Cr, and $^{66}$Zn 
\citep{meyeretal96, woosley97, nomotoetal97, brachwitzetal00, daveetal17, morietal18}.

A very low density region surrounding the white dwarf, sometimes referred to in
the literature as ``fluff,'' is required by Eulerian grid-based simulations,
which cannot treat empty space without some matter density.  The fluff is
chosen to have an initial density of $10^{-3}$ g cm$^{-3}$ and temperature of
$3 \times 10^7$ K, and is dynamically unimportant for the duration of the
models presented here. 


Since the deflagration energy release and the nucleosynthetic yield of $^{56}$Ni hinges 
critically on the bubble initial conditions, we investigate the earliest phases of the bubble 
evolution in Section \ref{sec:earlytime}.  The turbulent cascade behaves fundamentally differently in 2D and 3D,
and influences our choices in determining the spatial dimensionality of the 
simulations presented here.
In particular, in 2D, the turbulent cascade is inverse, proceeding from smaller
to larger scales \citep{kraichnan67}. In contrast, in 3D, the turbulent
cascade proceeds directly, from larger to smaller scales, where the energy is
dissipated at the smallest scales due to viscosity. In the early-time simulations, the bubble remains
laminar, and can be simulated in 2D. The fundamental distinctions between turbulence in 2D and 3D
have major ramifications for studying longer timescales, on which the flame becomes fully turbulent,
since physically-motivated flame-turbulence interaction subgrid models can {\it only} be realized in 3D.
Consequently, all longer-time simulations, in which the bubble enters a turbulent state
have been run in 3D with Cartesian geometry with a turbulence-flame
interaction model to capture enhanced burning on subgrid scales. All 3D simulations
were performed both with and without the turbulence-flame interaction (TFI) model (described below),
thereby spanning a range of possible outcomes on the flame propagation resulting from
unresolved turbulence. Test 2D simulations in cylindrical coordinates
led to unphysical behavior in the turbulent phase, including spurious
surface protuberances burning in the radial direction and thus significantly altering the
simulation outcomes in comparison to 3D models. Artificial outcomes were
particularly significant for runs with ignition points close to the center
of mass of the white dwarf, where unphysical burning in the radial direction
has the largest impact.


Our Cartesian domain extends from $-6.5536\times 10^{5}$~km to 
$+6.5536\times 10^{5}$~km in each direction, with a maximal 
refinement down to $\Delta=4$~km. 
We employ several refinement criteria, which are designed to follow the nuclear
burning of the models at high resolution, while also minimizing the resolution
in the very low density regions outside the white dwarf itself. 
Our simulations seek to maintain the highest resolution
in the burning region behind the flame surface, and employ a standard 
density gradient criterion to refine when the density gradient parameter
exceeds $0.1$, and derefines when it is beneath $0.0375$.
Further refinement criteria seek to derefine in the fluff 
and in regions outside of active burning, derefining one level if
the energy generation rate is lower than 
$5 \times 10^{17}$~erg~g$^{-1}$~cm$^{-3}$, and completely to level one 
if the density is below $10^3$~g~cm$^{-3}$.
Except for their resolution and threshold, these criteria are the same as
in \citet{townsleyetal09}.  Furthermore, because the ejected ash continues to
expand over time, the computational cost of following the ejected ash grows without bound.
Consequently, we impose an additional derefinement
outside a radius of $4000$~km to $\Delta=128$~km, which only impacts the ejected
mass. We increased this derefinement radius to $6000$~km for offsets $r_0\lesssim
20$~km, where the pre-expansion can reach similar radii. 

The single flame bubble's initial size is limited by the hydrodynamic resolution
of our simulations. At a resolution of $\Delta=4$~km for the flame front, we
assume an initial spherical shape with radius is set to $R_0=16$~km. 
In order for this to
be a reasonable assumption, a self-consistent evolution since appearance of
the hot spot should yield a negligible velocity profile and a self-similar
evolution preserving sphericality as discussed
in~\citet{VladimirovaModelflamesBoussinesq2007}.

The consistency of assuming a spherical ignition point can be assessed with
simple physical arguments. The flame polishing scale
$\lambda_\mathrm{fp}=4\pi S_l^2/(Ag)$, below which perturbations on the surface
are polished out \citep{timmeswoosley92}, implies that even if the initial
ignition was non-spherical it would become spherical soon afterwards. An explicit
numerical test confirming this was performed by~\cite{maloneetal14}.
Later perturbations to the sphericality
can arise from a turbulent background flow and the buoyant rise. The background
flow is small compared to the laminar flame speed of $\sim 100$~km/s, so that
sphericality should initially be sustained.
The impact of the buoyant rise on sphericality is closely linked to the question
of a negligible initial velocity field, whose amplitude however increases as the
bubble starts to rise. We assume the velocity field to become relevant when the
velocity from the gravitational acceleration $g$ reaches the order of the
laminar velocity, the stretching scale $l_{\rm fl}$. This should approximately
corresponds to $l_\mathrm{fl}=2S_l^2/(Ag)$ \citep{maloneetal14},
which primarily depends on the offset near the white dwarf's center. $A$
is the Atwood number of fuel and ash density. Note
that this criterion is stricter by a factor of $2\pi$ compared to the criterion
for sphericality due to perturbations: The flame bubble is expected to stretch
radially before the wrinkles in the flame front are not polished out anymore.
Alternatively to above estimator for $l_{\rm fl}$, we integrate the flame's
evolution based on \citep{fisherjumper15} to determine when the bubble's
velocity reaches the laminar flame speed, which is shown in
Table~\ref{tab:2Druns}. Only at small radial bubble
offsets $r_0$ from the center, which we are particularly interested in,
fulfill this condition. This length scales with $r_0^{-1}$ and therefore the
condition allows large bubbles at low offsets. As \citet{fisherjumper15}
argue that there is a critical offset at which the deflagration will burn through
the core, vastly changing the overall deflagration yield and thus its possible
detonation, a completely self-consistent evolution would be desirable at
these offsets. However, we are effectively limited by the required computational
resources and resolution. As an alternative for such self-consistent treatment,
we evolve 2D models for the linear phase, where deviations to 3D outcomes should
be negligible, which we can resolve sufficiently well.

We incorporate a turbulence-flame interaction model presented in \citet{jackson_power-law_2014}
implementing a specific model of power-law wrinkling based on that proposed by
\citet{charlette_power-law_2002}.  The reaction front is modeled by a
reaction-diffusion front which propagates with a speed based upon
the estimated physical features of the wrinkled physical flame whose width
is, for most of the interior of the WD, many orders of magnitude smaller than
the computational grid scale.  Due to the interaction of turbulence
with the flame, the location of the reaction front, as coarsened to a filter scale $\Delta$ consisting of a few grid cells, is approximated to propagate at a turbulent flame speed $s_t =
\Xi s_l$, where $s_l$ is the physical laminar flame speed, and $\Xi$ is called
the wrinkling factor.  The wrinkling is given by
\begin{equation}
\Xi = \left(1+\frac{\Delta}{\eta_c}\right)^{1/3},
\end{equation}
where $\eta_c$ is the cutoff scale for wrinkling, and is dependent upon
local properties of both the turbulence and the physical flame.  In this
model, $\eta_c$ is the inverse of the mean curvature of the flame surface,
and is determined by assuming equilibrium between subgrid flame surface
creation due to wrinkling by the turbulence and flame surface destruction by
flame surface propagation and diffusion.  This turbulence-flame interaction model leads to a turbulent flame speed $s_t$ that
is approximately equal to the characteristic speed of turbulent fluctuations on the filter scale, $u'_\Delta$ at intermediate densities,
$10^8$-$10^9$~g~cm$^{-3}$, as can be seen  in Figure~4 in \citet{jackson_power-law_2014}.  The 
turbulent flame speed falls off to the laminar flame speed at lower densities where the flame is too thick
and slow to support wrinkling.
At high densities where the flame is effectively polished by the high laminar speeds, the turbulent flame speed also approaches to the laminar flame speed value.
Performing the calculation of the cutoff scale for wrinkling, $\eta_c$, requires a measurement of the turbulence on the filter scale, $u'_\Delta$, and
makes the physical assumption that the subgrid turbulence is homogeneous,
isotropic, and follows Kolmogorov's theory on the filter scale. 
As shown by \citet{zingaleetal05}, buoyancy-driven turbulence becomes increasingly
homogeneous and isotropic on small scales, implying the last assumption is valid provided
the filter scale is sufficiently small.   

\begin{figure}
\begin{center}
\includegraphics[width=1.0\columnwidth]{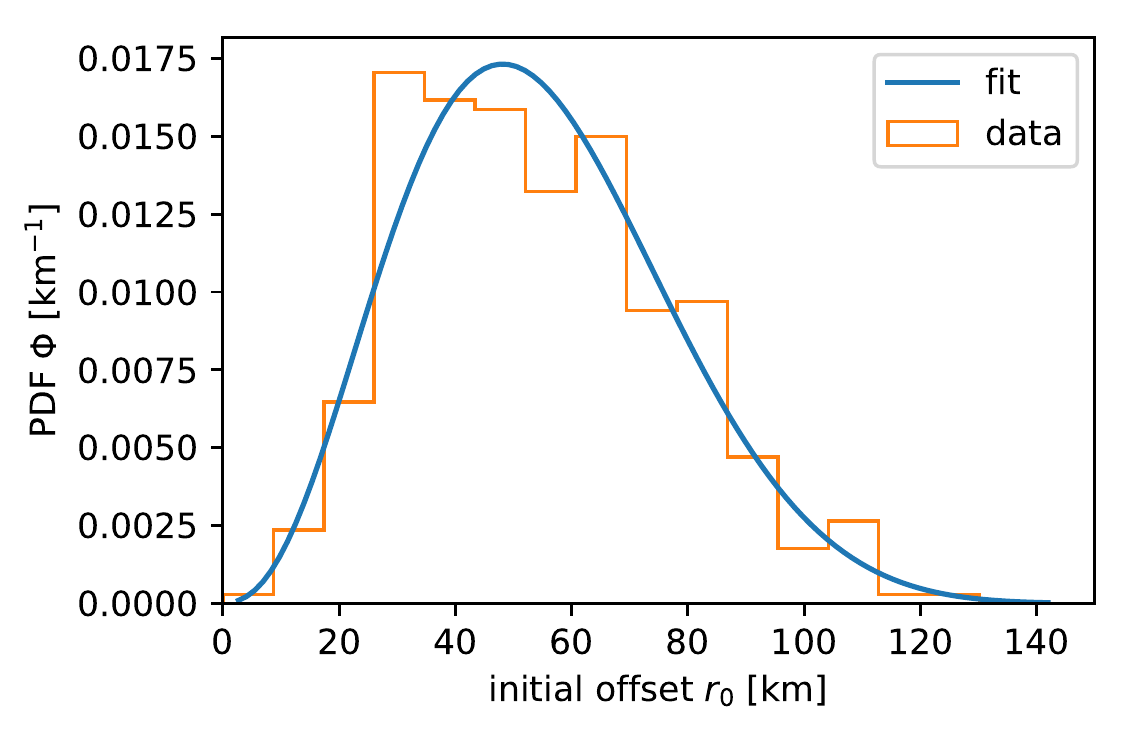}
\caption{Radial hot spot distribution and fit function for raw data used in~\citep{nonakaetal11}.}
\label{fig:hotspotPDF}
\end{center}
\end{figure}

From ~\citet{nonakaetal11} we obtain the probability density $P(r)$ of
hot spots forming at a certain distance from the center of mass.
Using the raw data and the same methodology, we create a histogram for such
distribution and a fit to this distribution, see Figure~\ref{fig:hotspotPDF}.
Shown is the probability density function per unit length.
Under the assumption that the probability density
per volume only mildly changes near the center of mass,
this implies a $P(r)dr\propto r^2dr$ scaling at low offsets due to the
shrinking volume available for hot spots to occur. 
While not exactly fulfilling this consideration, 
we obtain a reasonable fit using a $\beta$-distribution, obtaining
an expectation value of $\langle r_0\rangle=48$~km and a probability of 2.2\%
for hot spots forming at $r_0<16$~km, the critical ignition radius 
determined by \citet {fisherjumper15}.

We simulate the outcomes of varying ignition offsets for the given progenitor
according to the this probability distribution and choose a representative
range of initial offsets with an initial bubble radius of $R_0=16$~km shown in
Table~\ref{tab:3Druns}. We utilize both 2D and 3D simulations.
The size of the initial bubble is naturally limited by the
simulation resolution. As demonstrated analytically in \citep {fisherjumper15},
the flame bubble's dynamics vastly change at low initial offsets. We employ 2D
simulations to investigate the initial stages of the bubble dynamics at very
high resolution. Moving to 2D is a reasonable strategy in this case,
as we are only interested in the initial, linear phase.
In 2D, we employ a maximal resolution of $0.25$~km and
an initial bubble radius of $2$~km, at initial offsets ranging from $0-50$~km as
listed in Table~\ref{tab:2Druns}.

The 3D simulations are evolved until they can undergo a detonation as GCD.
The precise conditions under which a DDT may arise are still a matter of active
investigation, though recent three-dimensional simulations may shed
further light on this issue \citep {poludnenkoetal11, fisheretal18}.
We adopt conservative criteria for detonation initiation based upon studies of the
the Zel'dovich gradient mechanism \citep {Seitenzahl_2009}, which demonstrate
that the critical length above temperatures $\simeq 2 \times 10^9$ K at
a density $10^7$ g cm$^{-3}$ become of order 1 km, and a detonation is deemed likely.  
Further, in this paper, we evolve all 3D models within the context of the GCD scenario.
As discussed in the introduction, current {\it ab initio} calculations point
towards offset single-point
ignitions, which favor both the GCD and DDT scenario over pure deflagrations. 
Because the GCD model involves further evolution post-bubble breakout, 
it generally predicts a greater deflagration energy release than the DDT model, for
an otherwise identical WD progenitor and flame bubble ignition model. 
Consequently, consideration of the GCD model yields a lower limit for the 
mass of $^{56}$Ni due to a lower central density $\rho_c$ at the time of
detonation, in comparison to DDT models.

Artificial detonations can occur due to temperature oscillations arising  as
numerical artifacts from degenerate stellar equation of state coupled with
hydrodynamics close to
discontinuities~\citep{ZingalePiecewiseParabolicMethod2015}. These oscillations
are particularly striking during the flame bubble's buoyant rise. In order to
prevent detonations arising from these artifacts, we restrict detonations to
occur in the southern hemisphere ($z<0$~km).

\begin{table}
\caption {\label{tab:2Druns} Performed 2D runs with a maximal resolution
of $\Delta=0.25$~km and initial radius of $2$~km.}
\begin{ruledtabular}
\begin{tabular}{lll}
	Offset (km) & $\lambda_{\rm fp}$ (km) & $l_{\rm fl}$ (km)\\ \hline
	4  & $353.6$ & $40.9$ \\
  10 & $141.4$ & $27.2$ \\
	20 & $70.7$ & $17.6$ \\
	50 & $28.3$ & $8.2$ \\
\end{tabular}
\end{ruledtabular}
\begin{tabbing}
\end{tabbing}
\end{table}

\begin{table}
\caption {\label{tab:3Druns} Performed 3D runs with a maximal resolution
of $\Delta=4$~km.}
\begin{ruledtabular}
\begin{tabular}{llll}
	Offset (km) & TFI & 
	$t_{\rm det}$ (s) & M$_{\rm Ni56}$ (M$_\odot$) \\ \hline
 0   &  \cmark/\xmark   & failed$^{\rm a}$/failed$^{\rm a}$ & $0.56$/$0.35$ \\
 16   &  \cmark/\xmark  & $3.70$/$2.92$ & $1.08$/$1.05$ \\
 20   &  \cmark/\xmark  & $3.34$/$3.27$ & $1.12$/$1.06$ \\
 32   &  \cmark/\xmark   & $2.61$/$2.54$ & $1.14$/$1.14$ \\
 40   &  \cmark/\xmark   & $3.06$/$2.42$ & $1.09$/$1.13$ \\
 50   &  \cmark/\xmark   & $2.48$/$2.31$ & $1.14$/$1.20$ \\
 100   &  \cmark/\xmark  & $2.26$/$2.12$ & $1.21$/$1.20$ \\
 125   &  \cmark/\xmark   & $2.21$/$2.08$ &  $1.22$/$1.20$ \\
\end{tabular}
\end{ruledtabular}
\begin{tabbing}
$^{\rm a}$Model fails to detonate. \hspace{25pt} 
\end{tabbing}
\end{table}

\section{Results}
\label{sec:results}

We first discuss the early phase of 
bubble evolution. The early linear phase of evolution is similar in both 2D and 3D,
so we investigate the early linear evolution in high-resolution 2D models and compare these 
against semi-analytic predictions. 
We then move on to examine the subsequent nonlinear evolution through breakout
and detonation in full 3D, which we also evolve starting with the linear phase,
but at lower resolution.

\subsection {Early Linear Evolution in 2D} 
\label{sec:earlytime}

\begin{figure}
\begin{center}
\includegraphics[width=1.0\columnwidth]{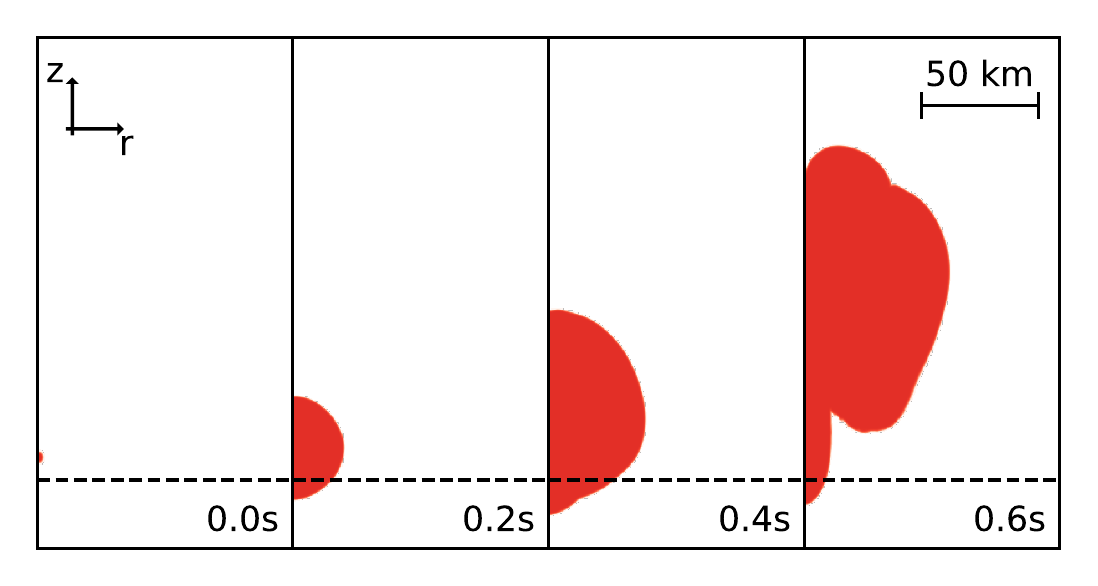}
\caption{Deflagration phase during the first $0.6$~s for a 2D model
	with $10$~km offset, $2$~km initial radius and a resolution of $0.25$~km. 
	The dashed line shows $z=0$~km. On the coordinate axis $r$ denotes $\sqrt{x^2+y^2}$.}
\label{fig:slices_linphase}
\end{center}
\end{figure}

\begin{figure}
\begin{center}
\includegraphics[width=1.0\columnwidth]{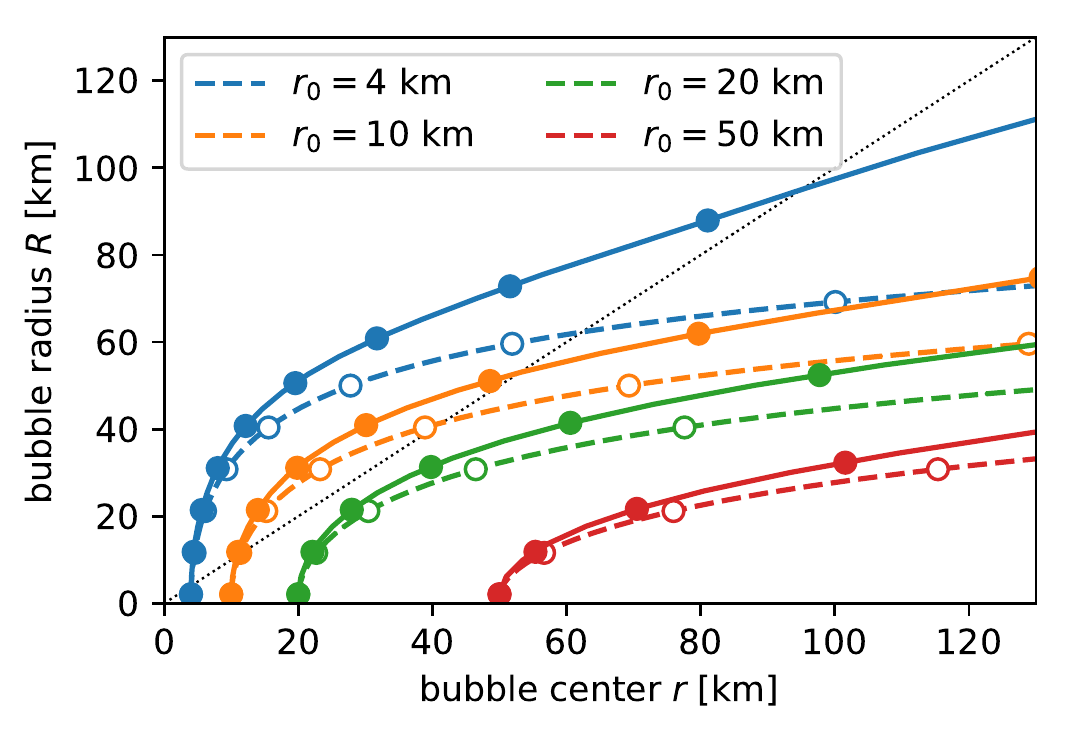}
\caption{Bubble evolutionary tracks shown for both 2D hydrodynamic simulations
  as well as for the analytic solution in ~\citep{fisherjumper15}. The plot
  shows the bubble radius $R$ versus offset radius $r$. The evolution of
  different initial offsets $r_0$ is shown as the solid curve for simulations
  and as the dashed curve for the analytical solution. The dots represent time
  steps of $0.1$~s, starting with $0.0$~s. States above the dotted line have
  burned through the white dwarf's center of mass.}
\label{fig:fj15_compare}
\end{center}
\end{figure}

\begin{figure}
\begin{center}
\includegraphics[width=1.0\columnwidth]{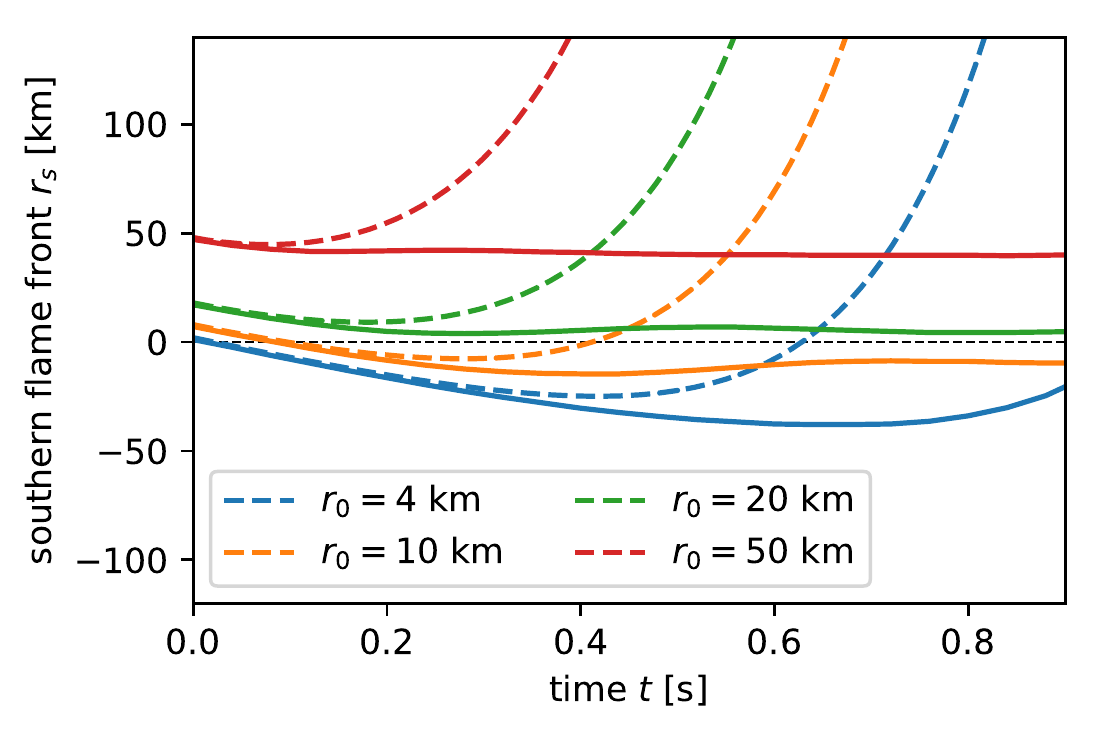}
\caption{Position of the southern flame front $r_s$ as a function of time $t$
after ignition. Line style and color are chosen as in Figure~\ref{fig:fj15_compare}.}
\label{fig:fj15_compare_southernfront}
\end{center}
\end{figure}

Figure~\ref{fig:slices_linphase} shows the slices of the flame bubble's 
evolution in its laminar phase for $r_0=10$~km. The burned material grows 
spherically as long as the buoyant velocity is small and the bubble's
size stays below the flame polishing scale. As the bubble grows, the acceleration
for material at the northern and southern flame front start to differ and the
bubble becomes elongated along the initial offset's direction until a
plume forms at the northern front.
Interestingly, the southern flame front's laminar speed seems to be
countered by the background flow from the buoyant rise at the northern front.

In Figure~\ref{fig:fj15_compare},
we show the evolution of the flame bubble's radius $R(r)$ as a function of the
bubble offset $r(t)$. For the simulation data,
the volume-equivalent spherical radius ($R=\sqrt[3]{3V/4\pi}$) deduced from the
burned volume $V$ is shown. We compare our results from the initial
phase of linear growth with the analytic model presented
in~\citep{fisherjumper15} and find them to be in good agreement for
the first tenths of seconds, particularly for larger initial offsets. The analytic description starts to fail as
the velocity from the buoyant rise becomes inhomogeneous across the bubble,
effectively stretching the bubble due to a lower speed
at the southern flame front. Figure~\ref{fig:fj15_compare_southernfront} 
shows the position of the southern flame front's position for the analytic
and numerical evolution, which start to differ as the analytic solution does
not incorporate an inhomogeneous velocity/acceleration field.
The resulting elongation gives rise to a stem being left behind the rising plume
at the northern front.
Even when the southern flame front crosses the center of
mass it will not buoyantly rise towards the opposite pole but is confined close
to the center of mass on the relevant time-scale due to the background
flow caused by the buoyant rise of the ash on the northern hemisphere.

\subsection {Non-linear Evolution and Detonation}

\begin{figure}
\begin{center}
\includegraphics[width=1.0\columnwidth]{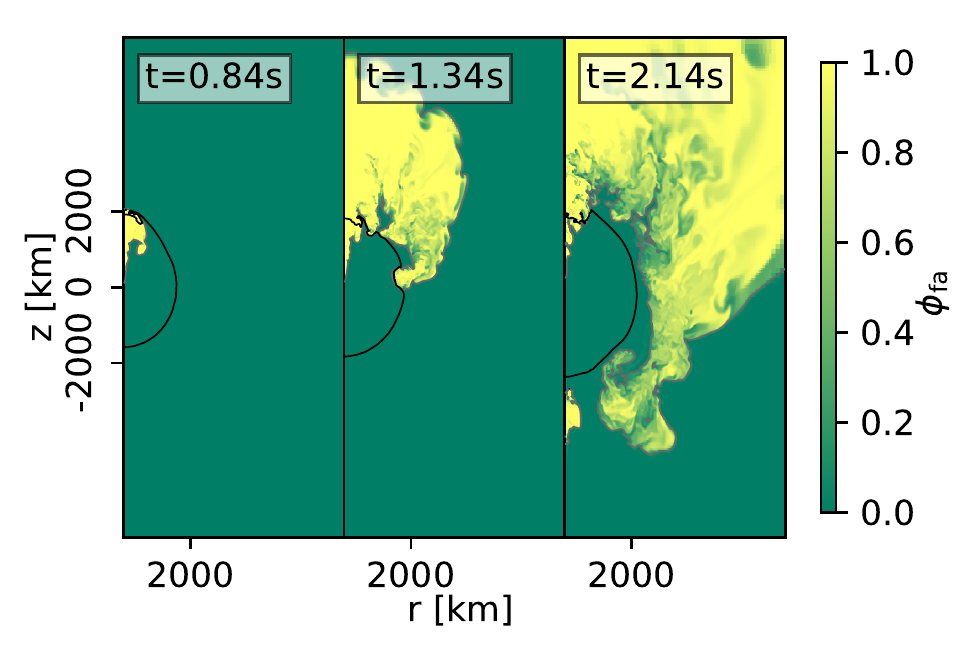}
\caption{Time series for run with enhanced burning at an offset of $100$~km
in three stages: Breakout, equator crossing and prior to detonation. Slices
are shown in the positive quadrant of the x-z plane with $y=0$~km with
$r=\sqrt{x^2+y^2}$. The colormap indicates the amount of burned material
$\phi_{fa}$. The solid line shows the density contour for $\rho=10^7$g cm$^{-3}$.}
\label{fig:slices_TFI_r100}
\end{center}
\end{figure}

\begin{figure}
\begin{center}
\includegraphics[width=1.0\columnwidth]{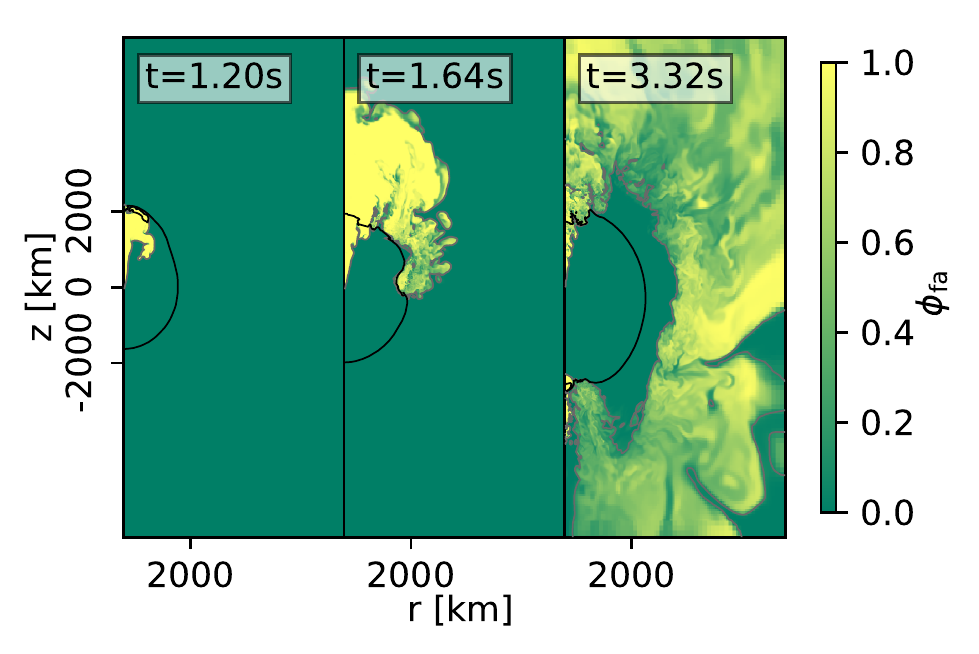}
\caption{Time series for run with enhanced burning at an offset of $20$~km
  analogous to Figure~\ref{fig:slices_TFI_r100}.}
\label{fig:slices_TFI_r20}
\end{center}
\end{figure}

\begin{figure}
\begin{center}
\includegraphics[width=1.0\columnwidth]{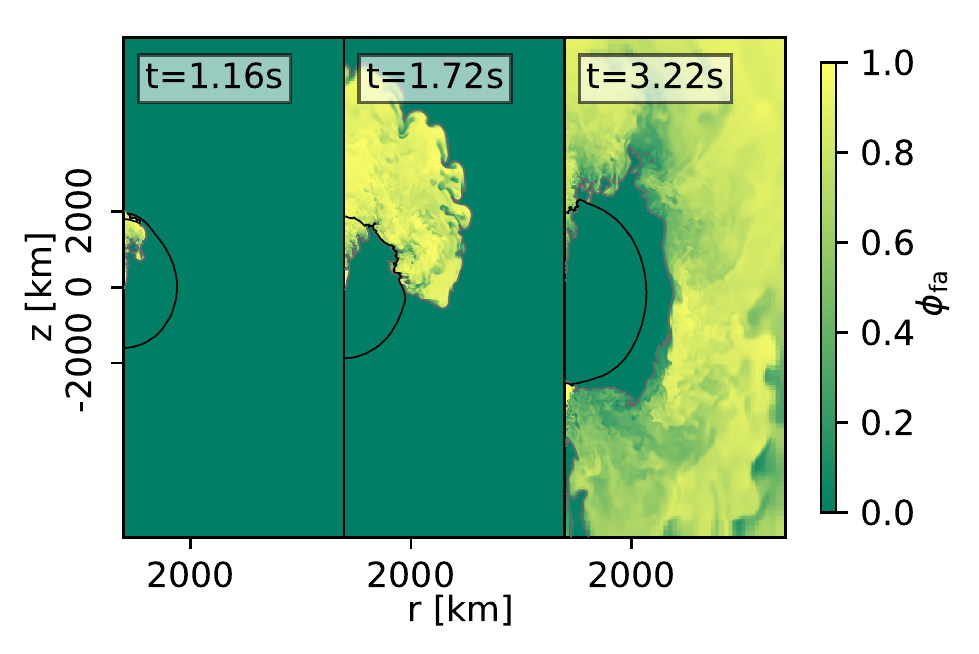}
\caption{Time series for run without enhanced burning at an offset of $20$~km
  analogous to Figure~\ref{fig:slices_TFI_r100}.}
\label{fig:slices_noTFI_r20}
\end{center}
\end{figure}

With formation of a rising plume the evolution becomes non-linear and depends
on the imposed flame model as presented in Section~\ref{sec:methodology}. To
capture the flame's turbulent rise, we evolve 3D models from ignition to
detonation for the parameters listed in Table~\ref{tab:3Druns}.
As we find the 3D runs to remain mostly symmetric, we show slices in the $z$-$x$
plane restricted by $x\geq 0$~km and $y=0$~km. 
Figure~\ref{fig:slices_TFI_r100} and~\ref{fig:slices_TFI_r20} show 
the evolution of the white dwarf for $20$~km and $100$~km offset.
Figure~\ref{fig:slices_noTFI_r20} also shows
the evolution of a $20$~km offset model, but without enhanced burning that the
prior two models use. Because the evolutionary
timescales for each run depend on the initial conditions chosen, the slices 
for each run are chosen with respect to the state of the flame, and not in
absolute time. In particular, in each plot, the first frame shows the breakout
of the flame at the star's surface. The next frame depicts the
post-breakout flame crossing the $z = 0$ equator on the star's surface.
The last frame shows the model just prior to detonation across from the point of
breakout.

While low offsets also have a slightly larger distance to the WD's surface, the
evolution of shown models demonstrate the delay to breakout in comparison to
larger offsets due to the smaller buoyant force near the center of mass,
increasing the breakout time by roughly $0.3$~s for $r_0=20$~km over
$r_0=100$~km. Smaller initial offsets, lead to a slightly increased plume size
both in radial and tangential direction with respect to the center of mass
across different initial offsets.

After breakout, the flame front travels around the white dwarf and
eventually reaches the point opposing the point of breakout. Material of the
envelope is pushed in front of this flame front into this opposing
point.  At larger offsets than $40$~km, such as the shown $100$~km run, the
ram pressure building up suffices to trigger a detonation before the ash reaches
the opposing point. For smaller offsets than roughly $40$~km, such as the runs with
an initial offset of $20$~km, the ram pressure is insufficient to trigger a
detonation upon the flame reaching the opposite pole and a detonation only
occurs after a subsequent partial recontraction of the white dwarf delaying the
detonation. For some offsets lower than the shown $20$~km, the white dwarf might
not detonate upon recontraction either.

The difference of the enhanced burning model seems to be only moderate for the
shown slices at offset $r_0=20$~km. The evolutionary phases represented by the
slices coincide between flame models, while without enhanced burning a larger
fraction of the material ejected from the white dwarf seems to be burned.

\begin{figure}
\begin{center}
    \includegraphics[width=1.0\columnwidth]{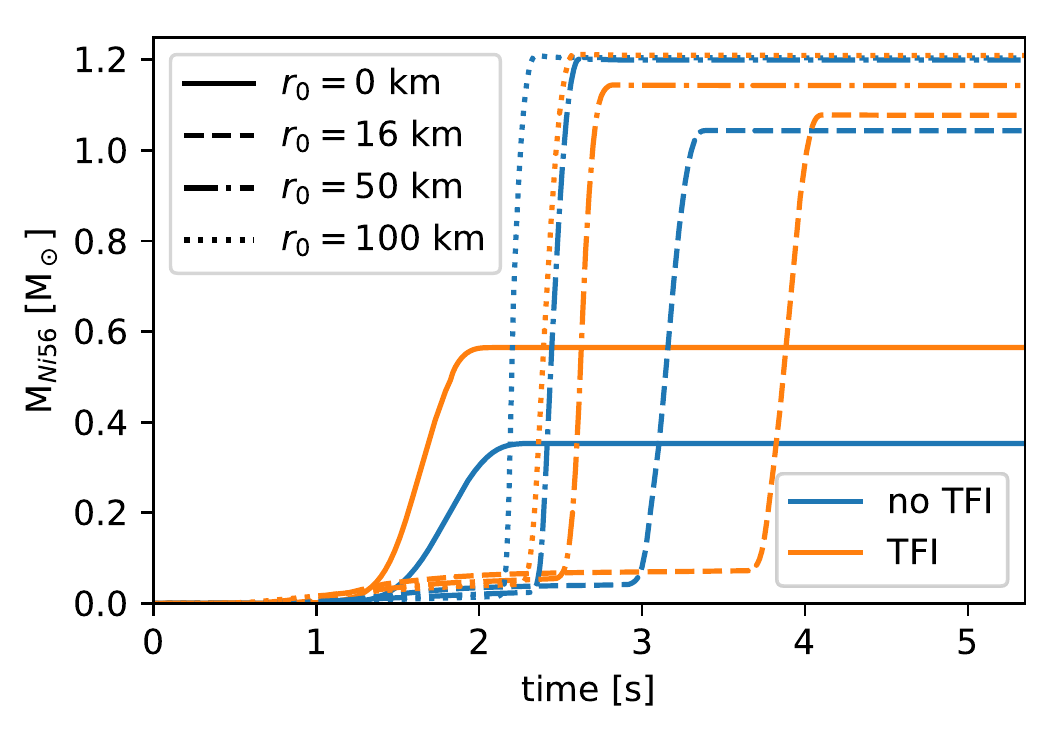}
\caption{Nickel 56 yield M$_{\rm Ni56}$ over time $t$ after ignition for the 3D
  simulations at selected initial offsets $r_0$ with and without TFI.}
\label{fig:Ni56_t3Dcombo}
\end{center}
\end{figure}

Figure~\ref{fig:Ni56_t3Dcombo} shows the estimated
$^{56}$Ni yield over time relative to the time of ignition. The $^{56}$Ni yield
in each model is obtained from the electron mass fraction $Y_e$, and by assuming
the  neutronization of IGE occurs in equal parts by mass of $^{54}$Fe and
$^{58}$Ni for all $Y_e$, which holds within 2\% in tabulated yields from
previous models \citep {meakinetal09, townsleyetal09}. The rapid increase of
M$_{\rm Ni56}$ occurring at $t\gtrsim 2.0$~s indicates the onset of detonation,
except for offset $r_0=0$~km, where a large growth sets in at $t=1.0$~s due to
turbulent deflagration.

We classify the progenitor's evolution into three different classes: failed, GCD
and PGCD. The PGCD scenario introduced by~\citep {jordanetal12a} shows a strong
recontraction phase ($2\cdot t_{det, GCD}\lesssim t_{\rm det,PGCD}$) due a
significantly increased deflagration yield from a many-bubble-ignition setup.
Given the smaller deflagration yields in our simulations due to the single
ignition hot spot, PGCDs here only show a mild recontraction (as e.g. indicated
by the evolution of the central density), making the
transition between PGCDs and GCDs gradual: Sometimes, there is no detonation
upon buildup of ram pressure but only when the fuel-ash mixture reaches the
southern pole even if no clear recontraction is present. We therefore do not impose a
binary criterion between those scenarios. 

\begin{figure}
\begin{center}
\includegraphics[width=1.0\columnwidth]{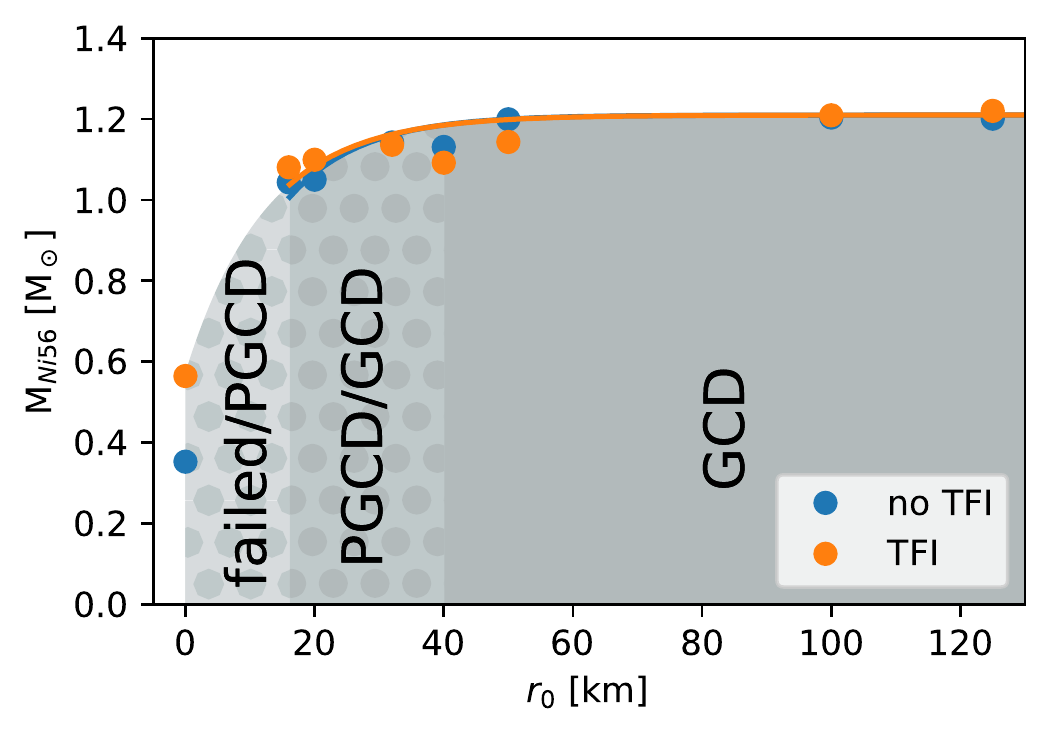}
\caption{Nickel 56 yields at different offsets in 3D with and without
enhanced burning including the $\arctan$-fit. Shaded regions mark offsets
resulting in GCD/PGCD/failed scenarios according to their label. Additionally
hatched regions mark transitions between scenarios due to computational
limitations and classification ambiguities (see text).}
\label{fig:Ni56_R}
\end{center}
\end{figure}

As Figure~\ref{fig:Ni56_R} and Table~\ref{tab:3Druns} show, the $^{56}$Ni yield
seems to converge towards roughly $1.21$~M$_\odot$ at large offsets $r\gtrapprox
100$~km with the progenitors undergoing the GCD scenario. 
As suspected by~\citep{fisherjumper15} the $^{56}$Ni yield decreases with lower
initial offset as visualized in Figure~\ref{fig:Ni56_R}. For runs $\lesssim
40$~km a transition towards PGCD-like scenarios takes place. The transition is
not monotonic, but
has a stochastic component whether the ram pressure from the initial
deflagration will suffice for an imminent detonation or not. For example, in our
simulations a $32$~km offset suffices for a GCD, despite the onset of a PGCD
already at $r_0=40$~km.

At some point we might
expect the PGCD scenario to fail. However, we lack the spatial resolution 
at very small radial offsets to determine the location of 
this transition. Thus, we are left with the artificial case of central ignition
for which the $^{56}$Ni yield from deflagration increases to
$0.56$~M$_\odot$/$0.35$~M$_\odot$ (TFI/no TFI),
depending on the flame model. However, there is no detonation so that the total
yield drops to these values as compared to higher initial offsets.
Due to the numerical expense of resolving the energy generating regions at
maximal resolution with AMR, we had to derefine to a maximal resolution of
$\Delta=8$~km at $t=2.00/1.82$~s for the no TFI/TFI scenarios with $r_0=0$~km.
Based on our parameter sampling, the transition
from PGCD to failed must occur at $0\textrm{ km}<r_0<16\textrm{ km}$ for the
chosen progenitor.

At large offsets, the yields with and without enhanced burning model vary only
marginally. However, at lower offsets the enhanced burning significantly adds
to the $^{56}$Ni yield, particularly for the failed events where differences
add up to $60\%$, but also for PGCD events in the order of up to $10\%$.

\subsection{Likelihood of $^{56}$Ni Yields}

We next compute the probability distribution of M$_{Ni56}$ outcomes for the
presented GCD SD channel models. The transformation from the hot spot
probability distribution $P(r_0)$ to the probability distribution
$P(M_{Ni56})$ of $^{56}$Ni outcomes is given as

\begin{equation}
\label{eq:transformation}
P (M_{\rm Ni56}) = \sum_{r_0 \in g^{-1} (M_{\rm Ni56})} \frac{P (r_0)}{\left|g' (r_0)\right|}.
\end{equation}
Here $g (r_0) \equiv M_{\rm Ni56}(r_0)$ is the amount of $^{56}$Ni 
produced as a function of offset radius $r_0$. 
$P(r_0)$ is the hot spot distribution found in \citep{zingaleetal11} shown
in Figure~\ref{fig:hotspotPDF}.
This relationship may be derived from Bayes' Theorem with minimal assumptions. We start with

\begin{align}
\label{eq:bayestheorem}
  P(M_{\rm Ni56}|r_0)P(r_0)=P(r_0|M_{\rm Ni56})P(M_{\rm Ni56}),
\end{align}
and assume a simplification that the $^{56}$Ni yield is solely 
determined by the offset position, leaving out possible uncertainties from
velocity flow and early bifurcations arising in the turbulent phase, one finds
\begin{align}
P(M_{\rm Ni56}|r_0)=\delta(g_\mathrm{M_{\rm Ni56}}(r_0)-M_{\rm Ni56}), \nonumber\\
P(r_0|M_{\rm Ni56})=\delta(r_0-g^{-1}(M_{\rm Ni56})), \nonumber
\end{align}
Here $\delta (x)$ is the Dirac delta distribution. Finally, using the identity

\begin{align}
  \delta(f(x))=\sum_i\frac{\delta(x-x_i)}{\left|f^{'}(x_i)\right|},
\end{align}
where we sum over the roots $x_i$ of $f(x)$,
we can rewrite equation~\ref{eq:bayestheorem} as equation~\eqref{eq:transformation}.

Even for a very similar stellar structure, we expect bifurcations arising from
the turbulent nature of the flame bubble's buoyant rise affecting the final
$^{56}$Ni yield. In our computations, we see a similar phenomenon from slight
offset changes and perturbations of the initial flame bubble. Therefore, the
yields we
obtained do not need to follow a monotonous relationship as only one possible
realization is drawn at a given offset. Given the numerical expense, neither can
we evaluate multiple runs at the same offset with slightly modified stellar
structure or flame bubble, nor can we afford to run more models at different
offsets. However, if deemed significant, additional parameters such as varied
background velocity field, could easily be incorporated by marginalizing over
such parameters. 
With our limited data sample we nevertheless try to obtain insights into the
resulting spread of $^{56}$Ni yields given the stochastic nature of the initial ignition
offset. To do so, we impose a strictly monotonous fit function for
M$_{\rm Ni56}$($r_0$) with an asymptotic yield at high offsets $r_0$ for which we use
\begin{align}
y(r_0) = \frac{y_{\rm max}+\Delta y}{2}+\frac{y_{\rm max}-\Delta y}{2}\cdot \tanh\left(s\cdot(r_0-r_s)  \right),
\end{align}
where $y_{\rm max}$ is the asymptotic yield at high offsets, $\Delta y$ the
spread between the two asymptotic branches, $r_s$ the position of the turning
point and $s$ characterizes the sensitivity of the $^{56}$Ni yield with respect
to the initial offset $r_0$. We fix the asymptotic yield to the approximate
value $y_{\rm max}$ found earlier. 

\begin{figure}
\begin{center}
\includegraphics[width=1.0\columnwidth]{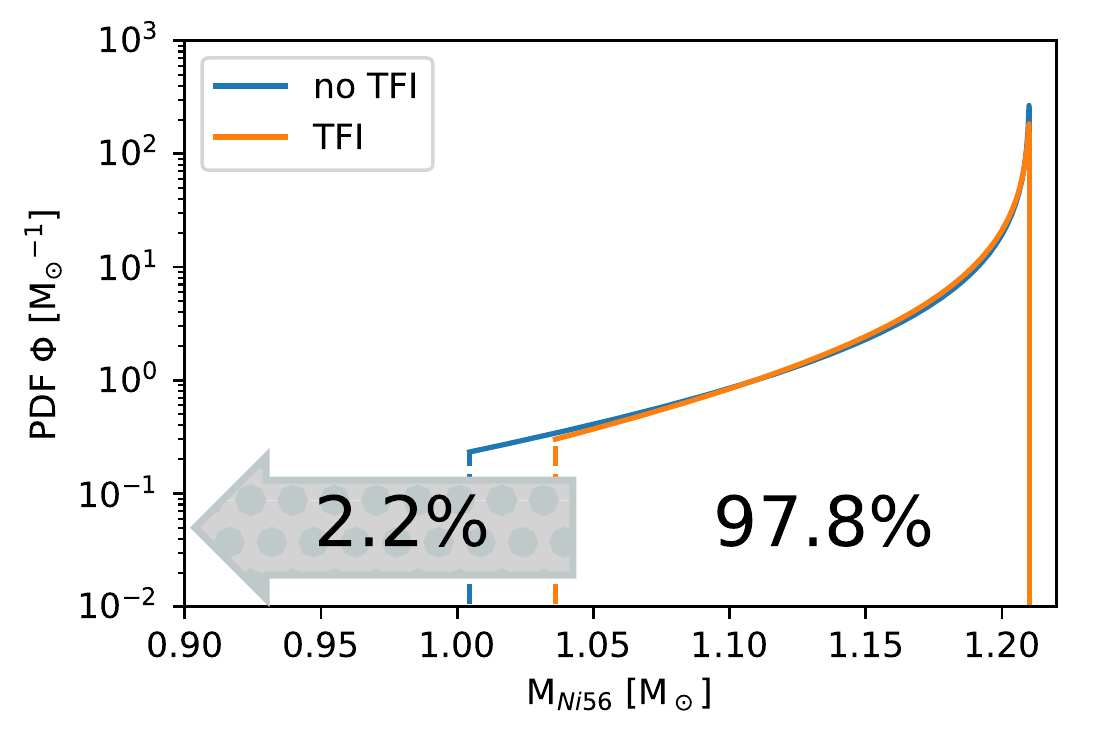}
\caption{Probability density function for Nickel 56 yields based on simulations
  and the hot spot probability density function.}
\label{fig:P_Ni56}
\end{center}
\end{figure}

The resulting distributions $P(M_{\rm Ni56})$
for our progenitor model with and without enhanced burning 
are shown in Figure~\ref{fig:P_Ni56}.
We show the probability distributions for events with offsets larger than
$16$~km, accounting for $97.8$\% of the ignitions.
The distribution shows a slightly larger spread for the non-TFI models due to
the lower $^{56}$Ni yield at low ignition offsets.
Nevertheless, we find that the majority of ignitions result in
a very confined $^{56}$Ni yield.

For given hot spot distribution $P(r_0)$ and the outcomes M$_{\rm Ni56}(r_0)$
of the simulated progenitor, we get a stochastic spread $P(M_{\rm Ni56})$
outcomes that is strongly favors overluminous events with a $^{56}$Ni yield of
$\sim 1.2$~M$_\odot$ with a standard deviation of $\sigma\sim 0.03$. However,
the total spread in outcomes of detonating models
$\delta=\max(M_{\rm Ni56}(r_0))-\min(M_{\rm Ni56}(r_0))$ due to the
stochasticity of hot spots forming is significantly larger. We are limited by
the hydrodynamic resolution, but find that $\delta \gtrsim 0.2$M$_\odot$.

We find $\sigma\ll\delta$ for both TFI and no TFI models as the $^{56}$Ni yield
is already close to asymptotic value of $1.21$M$_\odot$ at radii $r_0\sim
50$~km, which is the most likely point of ignition for the assumed hot spot
distribution. For a hot spot distribution peaking closer to the WD's center of
mass, the standard deviation could be significantly higher. The exact shape of
the left tail of the distribution is highly uncertain as it depends on the
chosen fit function given our sparse sampling. Similarly, we expect modeling
uncertainties, e.g. due to the lack of a velocity field, to propagate most
severely into the $^{56}$Ni yield and the resulting probability distribution at
low offset radii as the buoyant evolution can be strongly enhanced or delayed.
Other stochastic parameters, such as the state of the WD's velocity field, were
not considered, but would have to be marginalized over to obtain probability
distribution in a more elaborate study.

\section {Discussion}
\label{sec:discussion}

We have shown how the ignition offset probability distribution directly links to a range of
SNe Ia outcomes, parameterized by the $^{56}$Ni yield.  
This range of SNe Ia outcomes is intrinsically connected to the 
turbulent convective velocity field in the near-$M_{\rm Ch}$ WD progenitor, which
causes the ignition of the SD channel to be inherently stochastic and unpredictable.  
The physics of the SD channel is complex and is subject to numerous modeling uncertainties:
the pre-WD stellar evolution, accretion from the companion, possibly impacting the WD initial composition and structure, 
the physics of the simmering phase leading up to ignition, and the physics of turbulent
nuclear burning and detonation. 
In the following, we discuss this range of modeling uncertainties and to what extent each
of these effects may impact our conclusions.

Another crucial piece of physics underlying both the simmering phase and the nuclear
burning within the SNe Ia is the rate for C12 + C12 fusion. Recent experiments have 
measured this reaction rate for the first time for center-of-mass energies in the range
of 0.8 - 2.5 MeV, and demonstrated using the Trojan horse method, 
an enhancement in the cross sections by as much as a factor of 25 in a key temperature
range relevant to SNe Ia \citep {tuminoetal18}. While this work has been contested by other authors 
\citep {mukhamedzhanovetpang18} (and subsequently rebutted -- \citet {tuminoetal18b} ),
and will ultimately await additional confirmation,  it is important to recognize
the possible impact which uncertainties in this key reaction rate may have upon
the SNe Ia modeling. A higher reaction rate would increase the flame speed and
might particularly change the M$_{\rm Ni56}$ outcomes at low offsets where the buoyant
evolution is most sensitive to changes of our fiducial model.

In this work, we have incorporated the statistical distribution of ignition
points drawn from actual three-dimensional simulations of the convective
simmering phase of near-$M_{\rm Ch}$ WDs leading up to ignition. While this
approach has clear advantages over the majority of prior work, which typically
adopted ignition points in an arbitrary fashion, it is nonetheless still limited
by the fact that there has only been one high-quality three-dimensional
simmering phase simulation in a single WD progenitor completed to date. The
simulation has been performed at increasing resolution, and the distribution of
hot spot offsets appears to be converged \citep {zingaleetal11, nonakaetal11,
  maloneetal14}. However, it is conceivable that the distribution of hot spots
could be more centrally-condensed in WD progenitors with higher central density.

The mechanism underlying the initiation of detonation plays an important role in
SNe Ia theory, and much effort has focused upon whether the detonation mechanism
in a near-$M_{\rm Ch}$ SD scenario is a DDT or GCD \citep {ropkeetal07a,
  seitenzahletal16, daveetal17}. For example, because the DDT detonates prior to
bubble breakout, the stratification of the $^{56}$Ni and IGEs is generally more
centrally condensed, in broader agreement with observations of high M$_{\rm Ni56}$,
overluminous SNe Ia like 91T \citep {seitenzahletal16}. In the current work, we
have focused upon the GCD mechanism in our simulations in inferring the
intrinsic variation of the $^{56}$Ni production resulting from stochastic
ignition. However, if we were to have instead adopted a DDT criterion for
detonation initiation, the $^{56}$Ni distribution would be even more heavily
left-skewed. This is because, given identically the same WD progenitor and
ignition point, the DDT detonates prior to breakout, and consequently always
results in a less pre-expanded WD progenitor than a GCD \citep {daveetal17}. As
a result, the conclusion that the stochastic variance in $^{56}$Ni yields is
small, and the mean $^{56}$Ni yield is large, is not qualitatively modified
under the DDT scenario. 

In this work, we have begun with a quiescent WD, although the ignition arises in 
the WD interior, which is itself convective, and as a consequence of the transport of 
angular momentum from the accretion stream, may itself be rotating.
Indeed, recent work has shown that the effect of rotation may be significant enough
to weaken the convergence in the detonation region of a classical GCD \citep {garciasenzetal16},
although a PGCD might still be possible.
Furthermore, at low ignition offsets, the 
magnitude of the initial convective velocity field may have an impact on the
early flame bubble's evolution, and thus the $^{56}$Ni yield. As we start
our simulations with zero velocity, this adds an additional
uncertainty in the resulting M$_{\rm Ni56}$ distribution. On the one hand,
there is turbulence on small scales, distorting the flame front early on.
Expected velocities for this are small ($\sim 10$~km/s) 
with regard to the laminar flame speed ($\sim 100$~km/s), so that the flame
bubble's sphericality is still mostly unaffected until broken by its buoyant rise.
Minor shifts in a possible failed-to-detonated transition radius might be to be expected.
On the other hand, there is a possibility of the ignition point to occur
in a larger convective flow. If such hot spots form in convectively outward
moving regions as found by~\citep{nonakaetal11}, this will further decrease the
probability for ignitions that burn through the WD's center.
\citep{maloneetal14} ran a series of numerical simulations similar to our setup
for the deflagration phase, but additionally include a comparison of a setup with
self-consistent convective velocity field and one without any velocity field. 
In these simulations, the authors find the influence of the initial flow field
to increase as the initial ignition point is set closer to the center of mass,
particularly for an exactly centered ignition.

Stellar composition influences the final nucleosynthetic yield of a SD SNe Ia
through a variety of effects. The CNO metals of the WD stellar progenitor
ultimately yield  $^{22}$Ne during He burning.   
\citet{umedaetal99} suggested that a variation in the carbon 
abundance within the progenitor WD in the single-degenerate channel would 
impact the production of $^{56}$Ni. In particular, \citet{umedaetal99} 
conjectured that WDs with a richer C/O ratio would lead to a more turbulent 
flame, an earlier transition from deflagration to detonation at  higher 
densities, and hence a greater production of $^{56}$Ni.
 \citet {timmesetal03} demonstrated both analytically and numerically that the neutron excess carried
by $^{22}$Ne results in a decrease in the M$_{\rm Ni56}$ of the SN Ia event,
in direct proportion to the abundance of $^{22}$Ne. 
\citet {townsleyetal09} further considered a range  of additional compositional
effects influencing the final nucleosynthetic yields, including the ignition
density, the energy release, the flame speed, the WD structure, and the density
at which a possible deflagration-to-detonation transition
arises.  
The simulations with $^{22}$Ne mass fractions increasing from 0 to 0.02, which were run
long enough to determine a final $^{56}$Ni yield, demonstrate that the
combination of these effects result in a roughly 10\% decrease in M$_{\rm Ni56}$.
Similarly, we expect a slight decrease in $M_{\rm Ni56}$ based on complementary
work by \citet {jacksonetal10} investigating the impact of the
$^{22}$Ne content on the DDT density and the resulting $^{56}$Ni mass. 

Computational simulations of single-degenerate SNe Ia have subsequently explored
the influence of varying the C/O ratio within the progenitor WD in
the context of the DDT model \citep{kruegeretal10, ohlmannetal14}. 
These investigations have demonstrated that higher C/O ratios yield more
energetic and more luminous SNe Ia. 

Taken together, this body of work on SD SNe Ia generally supports the view that stellar
progenitor C/O ratio and metallicity play a role in determining the brightness
of a SN Ia event. However, at the same time, these models have demonstrated that
additional free parameters, including both the number and distribution of
ignition points, as well as the DDT transition density, have a combined effect
on the explosion energy comparable to that of the C/O ratio and stellar
progenitor metallicity.
Moreover, based upon this body of work, the combined influence of both a
decrease in the C/O ratio and an increase in the stellar progenitor metallicity
from the values assumed here (50/50 and 0, respectively), would result in a 10\%
- 20\% decrease in the $^{56}$Ni yields, which would quantitatively impact our
predicted M$_{\rm Ni56}$ distribution, but not alone yield a distribution more
closely resembling normal SNe Ia.

Most simulation models of near-$M_{\rm Ch}$ WDs adopt a central
density $\rho_c \simeq 2 \times 10^9$ g cm$^{-3}$, as we have in this
paper.  Because the electron capture rates are highly sensitive to the density,
higher-central density WDs generally produce greater amounts of stable 
IGE, and a lower $^{56}$Ni yield. Higher central density 
WDs significantly overproduce (relative to solar) a range of neutron-rich isotopes, including 
$^{50}$Ti, $^{54}$Cr, $^{58}$Fe, and $^{62}$Ni, and as a consequence,
were generally excluded from consideration as  near-$M_{\rm Ch}$ WD progenitors
\citep {meyeretal96, nomotoetal97, woosley97, brachwitzetal00}. However, if
SD near-$M_{\rm Ch}$ WDs constitute a small fraction of all SNe Ia, 
such high-central density WDs may not be rare occurrences. 
If the central density of the near-$M_{\rm Ch}$ WD is indeed higher than 
$\rho_c \simeq 2 \times 10^9$ g cm$^{-3}$, then the flame speed and the 
consequent deflagration energy release can be greater than considered here.
This can in turn lead to greater pre-expansion and a reduced amount of $^{56}$Ni
as shown in 2D simulations \citep{kruegeretal12,daveetal17},
possibly consistent with a normal or even a failed SNe Ia. However, the
qualitative outcome of an increased higher central density can vary as shown by
\citet{SeitenzahlTypeIasupernova2011}. 
In their 3D numerical study of the DDT scenario, the authors of the latter study
show that the central density is only a secondary parameter. However, their
study assumed multipoint ignition over a wide range of ignition kernels. When
single point ignitions are adopted, increased electron capture rates at higher
central densities lead to higher abundances of neutron-rich iron peak elements
at the expense of $^{56}$Ni \citep{daveetal17}.

\section{Conclusions}
\label{sec:conclusions}
In this paper, we investigated the impact of a single initial ignition's offset
$r_0$ for a single ignition point of a deflagration flame bubble in a fiducial
50/50 C/O WD with a central density of $2.2\times 10^{9}$~g~cm$^{-3}$ and an
adiabatic temperature profile leading up to Type Ia supernovae in the GCD
scenario.

We showed that a transition to failed SNe Ia (i.e. those events lacking a
GCD) occurs as $r_0$ falls below some offset below $16$~km.
Even for those white dwarfs detonating, the $^{56}$Ni yield spawns a range
of outcomes changing by $10-20$\% with a decreasing yield as $r_0$ approach the
radius where no detonation is triggered. 

Summarizing our key conclusions:

\begin {enumerate}

\item Stochastic range of outcomes. For chosen progenitor this corresponds to
  a spread of $\delta\gtrsim0.2$~M$_\odot$ for detonating models, even though
  the M$_{\rm Ni56}$ distribution is strongly left-skewed so that low M$_{\rm Ni56}$ are
  unlikely for the given probability distribution. This range of outcomes is
  stochastic and will add onto other variations from the different progenitors'
  stellar structure and evolution.

\item For non-centered ignitions, all ignitions lead up to an overluminous SNe Ia.
  We do not find a viable scenario from a single bubble ignition leading to a
  normal Type Ia for the progenitor used here, which is also commonly referenced
  in literature. This disfavors single degenerate progenitors as a contributing
  channel to failed and normal type Ia SNe. If this channel was to contribute to
  failed and normal type Ia supernovae, this would require readjustment and
  better understanding of the stellar structure and evolution, and flame
  dynamics.

\item (Quasi-)symmetric deflagrations around the center of mass, as commonly
  used in numerical studies, are most likely artificial constructs: Ignitions very close to
  the center are rare as shown by \citet{nonakaetal11} and even if such
  events occur, a strong asymmetry evolves as a background flow in direction of
  the outermost flame front counteracts the burning into other directions as
  numerically demonstrated here for offsets as small as 4~km. However, future
  in-depth studies
  on the likelihood of multi-ignition occurrences and their correlations with
  the turbulent velocity field might leave room for rare occurrences of
  symmetric deflagrations.

\end {enumerate}

{\bf Acknowledgements}  The authors thank Mike Zingale for generously providing the 
simulation data from his group's convective simmering phase models, which was used 
in Figure~\ref{fig:hotspotPDF}. The authors thank Pranav Dave and Rahul Kashyap for
insightful conversations. RTF also thanks the Institute for Theory and
Computation at the Harvard-Smithsonian Center for Astrophysics for visiting
support during which a portion of this work was undertaken. CB acknowledges
support from the Deutscher Akademischer Austauschdienst (DAAD). 
RTF acknowledges support from NASA ATP award 80NSSC18K1013. This work
used the Extreme Science and Engineering Discovery Environment (XSEDE) Stampede
2 supercomputer at the University of Texas at Austin's Texas Advanced Computing
Center  through allocation TG-AST100038. XSEDE is supported by National Science
Foundation grant number ACI-1548562 \citep {townsetal14}.

We use a modified version of the FLASH code 4.3 \citep{Fryxell_2000}, which was in part developed by
the DOE NNSA-ASC OASCR Flash Center at the University of Chicago, for our
simulations including the SN Ia modeling presented in~\citet{townsleyetal16}.
The authors have made use of Frank Timmes' hot white dwarf progenitor
(\burl{http://cococubed.asu.edu/code\_pages/adiabatic\_white\_dwarf.shtml}).
For our analysis, we acknowledge use of the Python programming language
\citep{VanRossum1991}, the use of the Numpy \citep{VanDerWalt2011}, IPython
\citep{Perez2007}, and Matplotlib \citep{Hunter2007} packages. Our analysis and
plots strongly benefited from the use of the yt package \citep{ytproject}.

\bibliography{converted_to_latex}
\end{document}